\documentclass[epj]{webofc}
\usepackage[varg]{txfonts}   % Web of Conferences font
\woctitle{The Time Machine Factory}
\begin{document}
\title{Cosmology and time}

\author{Amedeo Balbi\inst{1}\fnsep\thanks{\email{balbi@roma2.infn.it}}}

\institute{Dipartimento di Fisica, Universit\`a di Roma `Tor Vergata'\\
Via della Ricerca Scientifica 1, 00133 Roma, Italy.
}

\abstract{%
Time has always played a crucial role in cosmology. I review some of the aspects of the present cosmological model which are more directly related to time, such as:  the definition of a cosmic time; the existence of typical timescales and epochs in an expanding universe; the problem of the initial singularity and the origin of time; the cosmological arrow of time.}
\maketitle
\section{Introduction}
\label{intro}
Looking back to the history of cosmology, one can find a number of instances where arguments related to time and causality were used as a guide to speculations on the physical nature of our universe, starting from Newton's conditions for obtaining a stable matter configuration, to the solution of Olbers' paradox, to Einstein's introduction of the cosmological constant to obtain a static cosmological model, to the  solution to the oldness problem provided by inflationary scenarios. 

Similarly, it can be argued that many open problems in cosmology have something to do with time. In this contribution, I will try to give some (admittedly non-exhaustive) idea of such problems. 

\section{Time in standard cosmology}
\label{sec-1}

\subsection{Cosmic time}
The universe is a time machine. Every information an observer can gather on physical reality lives on her past light cone---the 2+1 surface connected to her by causal signals. Causality issues are then crucial in any discussion of cosmology, not only because they limit the amount of information we can gather, but also because they affect the physical mechanisms taking place in the universe, such as, for example, the formation of cosmic structures.

Although the division into past, present and future is observer-dependent, in our universe it is possible to define a universal cosmic time: this is a feature which is not at all obvious \textit{a priori}, and descends from special symmetry properties of spacetime on large scales. To be more specific, the geometry of our universe is aptly described in terms of the Friedman-Lema\^itre-Robertson-Walker (FLRW) 
\begin{equation}
ds^2=c^2dt^2-a^2(t)\left[\frac{dr^2}{1-kr^2}+r^2(d\theta^2+\sin^2\theta d\phi^2\right]
\end{equation}
which in turn is built upon two principles, namely 1) the Weyl's postulate requiring that fluid elements move on non-intersecting geodesics orthogonal to spacelike hypersurfaces and 2) the cosmological principle requiring that such spacelike hypersurfaces are maximally symmetric. Under this conditions, there exists a cosmic time \cite{rugh}, i.e. the proper time of comoving observers, such that 1) every comoving observer records the same cosmic history, and 2) at every time, the universe appears the same in every direction to every comoving observer. Then, we can unambiguously measure time and/or synchronize observations by recording the average density of matter in the universe or the temperature of the cosmic microwave background (CMB) radiation.

\subsection{Causality}
The structure of spacetime described by the FLRW metric is a geometrical construction, i.e., no assumption on the physical content of the universe is directly involved in its derivation. Physics enters in the time dependence of the scale factor, $a(t)$, which evolves according to Friedman equation:
\begin{equation}
H\equiv\left(\frac{\dot a}{a}\right)^2=\frac{8\pi G}{3}\rho-\frac{kc^2}{a^2}
\end{equation}
where $\rho$ is the average total density of the universe and $k$ is the constant describing the curvature (positive, negative or null) 
of the spatial sections of the FLRW metric. We actually observe the universe expanding according to Hubble's law ($v=Hd$, where $v$ is the velocity between any two points in space and $d$ is their separation), changing physical status accordingly. The universe evolves. The extrapolation of $a(t)$ at early times also (naively) suggests a definite origin of time at $a=0$, but this is not necessarily the case. In fact, when we extrapolate the status of the universe to early times, we get into a regime were known physics breaks down and we cannot draw any definite conclusion. Time may or may not have had an origin, and the universe may or may not have existed forever. However, in big bang models with standard matter and radiation, at any particular time we only observe a finite region of the universe, defined by the cosmic horizon $d_H(t)=c\int_0^t dt'/a(t')$. Our comoving observable volume evolved for a finite time: this is actually consistent with causality arguments, such as the solution to Olbers' paradox, although we cannot exclude that whatever spacetime exists outside our horizon existed forever.

In reality, the farthest we can look back is not the horizon, but the last scattering surface (LSS) where CMB photons last interacted with charged matter, before it recombined into neutral hydrogen atoms. CMB radiation has an incredibly precise blackbody energy distribution---a distinct signature that matter and radiation were in exquisite thermodynamic equilibrium during the first 380 000 years after the big bang. Furthermore, the intensity of CMB radiation shows a remarkable isotropy (one part in $10^5$), a feature which was for a long time considered a puzzling causality issue, since the observed CMB sky encompasses a large number of regions that were causally disconnected at the epoch of last scattering. There simply was no time between the big bang and last scattering to smooth the universe  through sub-luminal physical interaction. 

\subsection{Timescales}
An expanding universe (with conserved matter/energy per comoving volume) has a characteristic timescale, defined by the inverse of the Hubble parameter $H$, i.e. the Hubble time: 
\begin{equation}t_H\equiv H^{-1}=a/\dot a.
\end{equation}
This timescale goes as $t_H\sim (G \rho) ^{-1/2}$ in the big bang model (neglecting curvature), so that cosmic time effectively runs at a different pace at different epochs. An important consequence of this fact is that different physical processes go out of equilibrium as cosmic history unfolds, since their typical timescales eventually become longer than the Hubble time. Thus, time has a profound meaning in the big bang model, as opposed to models with a constant expansion rate, such as the now discredited steady-state scenarios.  Hubble time also sets the timescale for the age of the universe: with the measured value of the Hubble parameter at present, $H_0=74.3\pm 2.1$ km/s/Mpc \cite{freedman}, one gets $t_H\approx 13.7$ Gy. The actual age of the universe is found by integrating over the entire history of the universe as $t_0=\int da/\dot a$. With the current best fit cosmological parameters, one gets $t_0=13.74\pm 0.11$ Gy \cite{hinshaw}. Although in the big bang model it is to be expected that $t_0$ is roughly of the same order of $t_H$, the fact that in the currently accepted concordance model one has at present $t_H \simeq t_0$ is no more than a coincidence.  The Hubble time does not seem to be a fundamental timescale (i.e. it cannot be built upon a combination of physical constants) and the age of the universe is related to its content and expansion rate through a dynamical equation. There does not seem to be any fundamental reason, then, for the fact that its current value is huge when expressed in natural units, or $t_H\simeq t_0 \simeq 10^{60} t_P$, where $t_P=\sqrt{\hbar G/c^5}\simeq 5.4 \times 10^{-44}$ s is the Planck time. Is this just an initial condition feature? This is an interesting question since, unlike other physical systems, initial conditions for the universe might reveal some deep implication for fundamental models. It has also been pointed out \cite{barrowtipler} that the typical timescale of a main sequence star, $t_\star\sim (Gm^2_N/hc)^{-1}\times h/m_Nc^2\sim 10^9$ years, is very close to the present value of $t_H$. This fact was used early on \cite{rees} to argue against steady-state models, since in such models it would have no explanation, while in the big bang model it arises naturally: intelligent observers (if any) should first emerge precisely at the epoch when $t_H\sim t_\star$.

\subsection{Naturalness}
Naturalness problems with timescales go back at least as far as Newton. In Newton's cosmology, matter was spread uniformly over an infinite extension, something which required incredibly fine-tuned initial conditions (in a letter to Bentley, Newton compares the position of stars in his models to ``needles standing on their points'') \cite{harrison}. In modern terminology, Newton's model was unstable to perturbations on timescales $t\sim (G\rho)^{-1/2}$ (for reasonable choices of star distribution this time is a few tens million years) \cite{harrison}. Even the first general-relativistic model of the universe had problems with timescales. In his 1917 paper \cite{einstein}, Einstein introduced a cosmological constant $\Lambda_E=4\pi G \rho_{matter}$ in order to make the universe static, but, even so, the model was unstable to small perturbations: Einstein's universe could not remain static and/or homogeneous on timescales much larger than $t \sim 10^{10}$ y. Shortly after, de Sitter \cite{desitter} introduced his model with a non-zero $\Lambda$ and negligible matter content, expanding exponentially with a constant rate $H=\sqrt{\Lambda/3}$. With a ``natural'' value for the cosmological constant,  $\Lambda \sim l_P^{-2}$ (where $l_P=ct_P$ is the Planck length), the expansion timescale of the de Sitter model is $t_H\sim t_P$ which, again, gives a 60 order of magnitude mismatch with observations: in other words, $\Lambda$ has to be much smaller ($10^{-120}$  times) than one might expect. This is even more puzzling since one actually needs to have a non-zero $\Lambda$ (albeit a ridiculously small one) in order to fit cosmological observations. Note, in particular, that a cosmological constant was strongly suggested by the need to obtain a Hubble time compatible with lower limits on the age of the universe estimated from the oldest white dwarfs and globular clusters \cite{kraussturner} even before the discovery of accelerated expansion of the universe by type Ia supernovae data \cite{riess, perlmutter}.  

Similar ``age'' problems are found when one examines the total energy density of the universe: this has to be within a very narrow range around the critical value $\rho_c=3 H^2/8\pi G\sim 10^{-29}$ g/cm$^3$ to match the observed age of the universe, since models with densities larger or smaller than $\rho_c$ collapse or expand too rapidly. Indeed, cosmological observations show that $\rho$ is remarkably close to $\rho_c$ at present. But the $\rho = \rho_c$ condition is a highly unstable one (Newton's metaphor of ``needles standing on their points'' is particullarly apt for this situation too): in matter or radiation dominated universes, the energy density tends to rapidly diverge from $\rho_c$, so that $\rho$ had to be set to incredible precision at early times in order for the universe to last till its present age. This ``oldness'' problem has been known for a long time and it has been addressed by postulating an early de Sitter phase of accelerated expansion known as inflation \cite{guth}, which draws the energy density very close to $\rho_c$, thus setting the initial conditions for the following evolution of the universe. Inflation also solves the previously mentioned causality issue related to the extreme uniformity of the CMB, while generating density perturbations at early times, with the right properties to explain structure formation by gravitational instability in the time available from recombination to the present. Furthermore, it circunvents the issue of the origin of time (or of the singularity $a=0$), since the universe could have been very well existed forever before inflation took place \cite{aguirregratton}.

The problem with the cosmological constant, on the other hand, remains opened. In addition to the above mentioned 60 order of magnitude mismatch between the expansion timescale and the Planck time, there is a so-called ``coincidence'' or ``why now'' problem---namely, the fact that the cosmological constant contribution to the energy density today is roughly the same as that of matter \cite{hinshaw}, although the two were wildly different in the past (since the cosmological constant density remains constant throughout the evolution of the universe, while matter density decreases as $a^{-3}$). In fact, for most of the history of the universe no observation could have shown any clue that the cosmological constant was non-zero, and its contribution started to dominate only recently \cite{santos}. On a related note, once the cosmological constant becomes larger than the matter density it remains so forever (i.e. the universe approaches a future de Sitter phase), and it has been shown that, in the distant future of the universe, cosmology would become basically impossible \cite{krauss}. So we seem to live in a very special time in the history of the universe, a fact that looks puzzling until we realize that the timescale for the emergence of observers in an expanding universe is related to ``environmental'' prerequisites which in turn are related to the value of fundamental constants. It may then seem almost inescapable to resort to anthropic arguments \cite{weinberg} in order to explain the observed value of the cosmological constant (and perhaps of other physical constants). 

On the other hand, proposals have been made in order to explain the observed accelerated expansion of the universe without resorting to a non-zero cosmological constant. For example, it has been suggested \cite{wiltshire} that the accelerated expansion might in fact be an illusion due to the fact that we live in an overdense region of a very inhomogeneous universe, so that FLRW metric cannot be applied: there would be no universal cosmic time, and accumulating clock rate difference between different regions of the universe would lead to a misinterpretation of supernovae data. The definition of cosmic time might then play a crucial role in one of physics biggest misteries. It is worth noticing that one of the tools we might use to sort out such issues involves observations made on two null-cones separated by a few decades, an ambitious research program dubbed ``real-time cosmology'' \cite{realtime}: measuring variations of the cosmological redshift (of order 1 cm/s per year, within reach of high accuracy spectrographs coupled to future 50-100 m telescopes) could shed some new light, probing the dynamic of the expansions directly.

\section{Cosmology and the arrow of time}
Inflation solves some of the initial condition problems by ``preparing'' the universe in a state that evolves toward a standard big bang evolution starting from a ``generic'' state. How generic? This question might play a role in one of the biggest puzzles concerning the nature of time, i.e. where its apparent past-future directionality -- or ``arrow of time'' -- comes from (see \cite{davies} for a thorough discussion of this topic; a nice recent non-technical review is \cite{carroll}).  

\subsection{Entropy and the direction of time}
\begin{figure}
\centering
\includegraphics[width=0.8\textwidth]{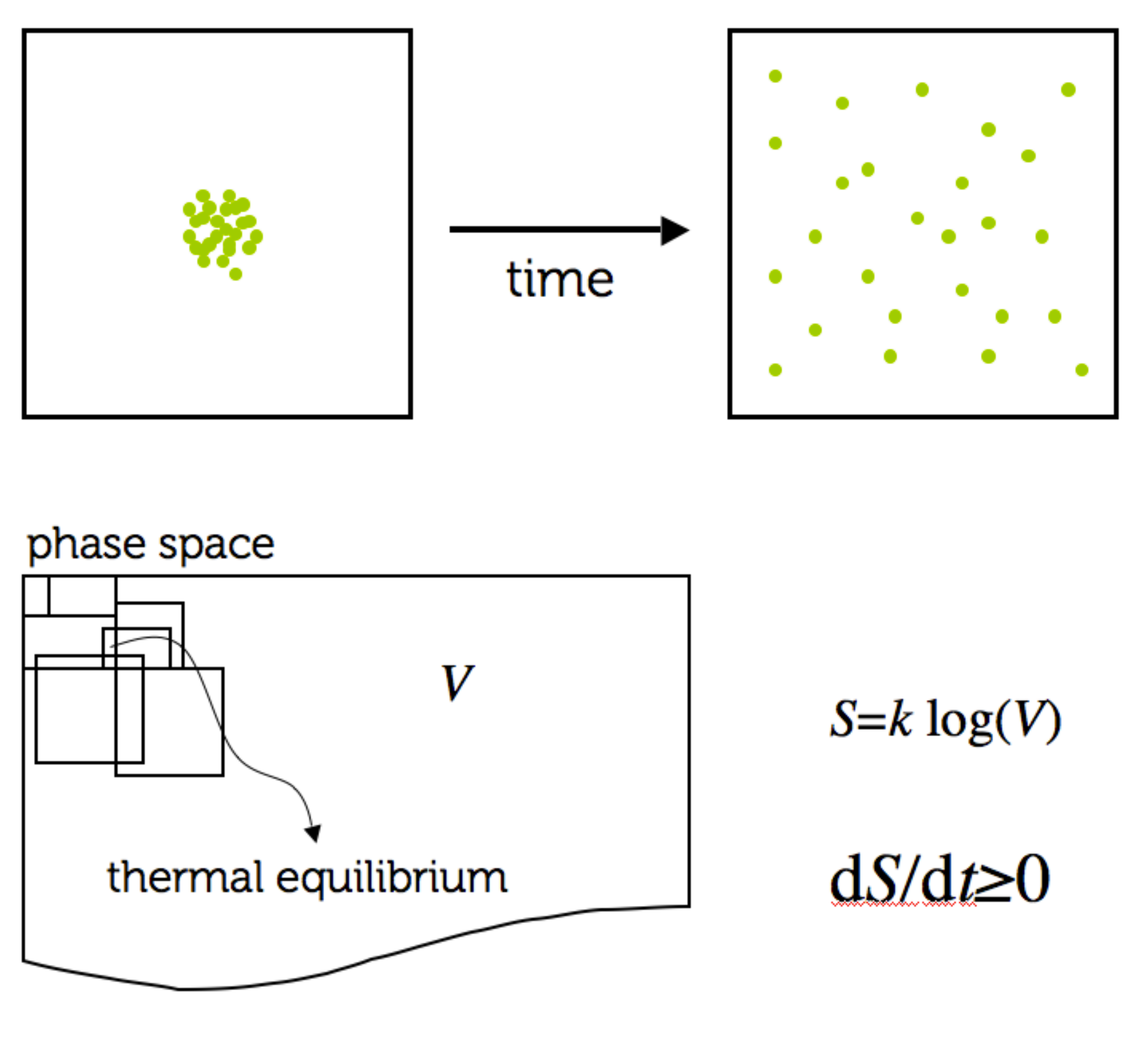}
\caption{The behaviour of an isolated system with many microscopical degrees of freedom can give rise to a macroscopical time asymmetry: if the initial state is prepared in such a way to occupy a small volume in phase-space, the subsequent dynamical evolution will bring the system in a state of macroscopical equilibrium. This is to be expected statistically, since the corresponding phase-space volume is much larger than the initial one. The reversal situation is never observed in everyday life because of its much lower probability.}
\label{fig-1}       
\end{figure}

We know that the flow of time in one particular direction (that from ``past'' to ``future'') is not written in the laws of nature, which are time-symmetric, but rather in the collective dynamical behaviour of  systems with many microscopical degrees of freedom, once special initial conditions are chosen and the system status is evaluated macroscopically through some coarse-graining. Then, the second law of thermodynamics ties the impression of a flow of time (and of irreversibility) to the increase of entropy, $S=k\log (V)$, where $V$ is the microscopical phase-space volume corresponding to a given macroscopical configuration. The standard example of this behaviour is that of a box containing a large number of particles (gas) which is initially prepared in a very unlikely state (e.g., with all particles clustered in a corner of the box): such a system will naturally tend to evolve from order to disorder, in the direction of increasing entropy, simply because there is a larger phase-space volume  to fill (Fig.~\ref{fig-1}). Note that the definitions of ``likely'' or ``unlikely'' or ``ordered'' or ``disordered'' only make sense after establising a coarse-graining procedure, i.e. counting the number of microscopical states that correspond to the same macroscopical situation. In this sense, the maximum entropy state, or equilibrium, of the system, is that corresponding to the unique state with particles uniformly spread in the container, although this situation can correspond to many different microscopical states for the particles. 

\begin{figure}
\centering
\includegraphics[width=0.8\textwidth]{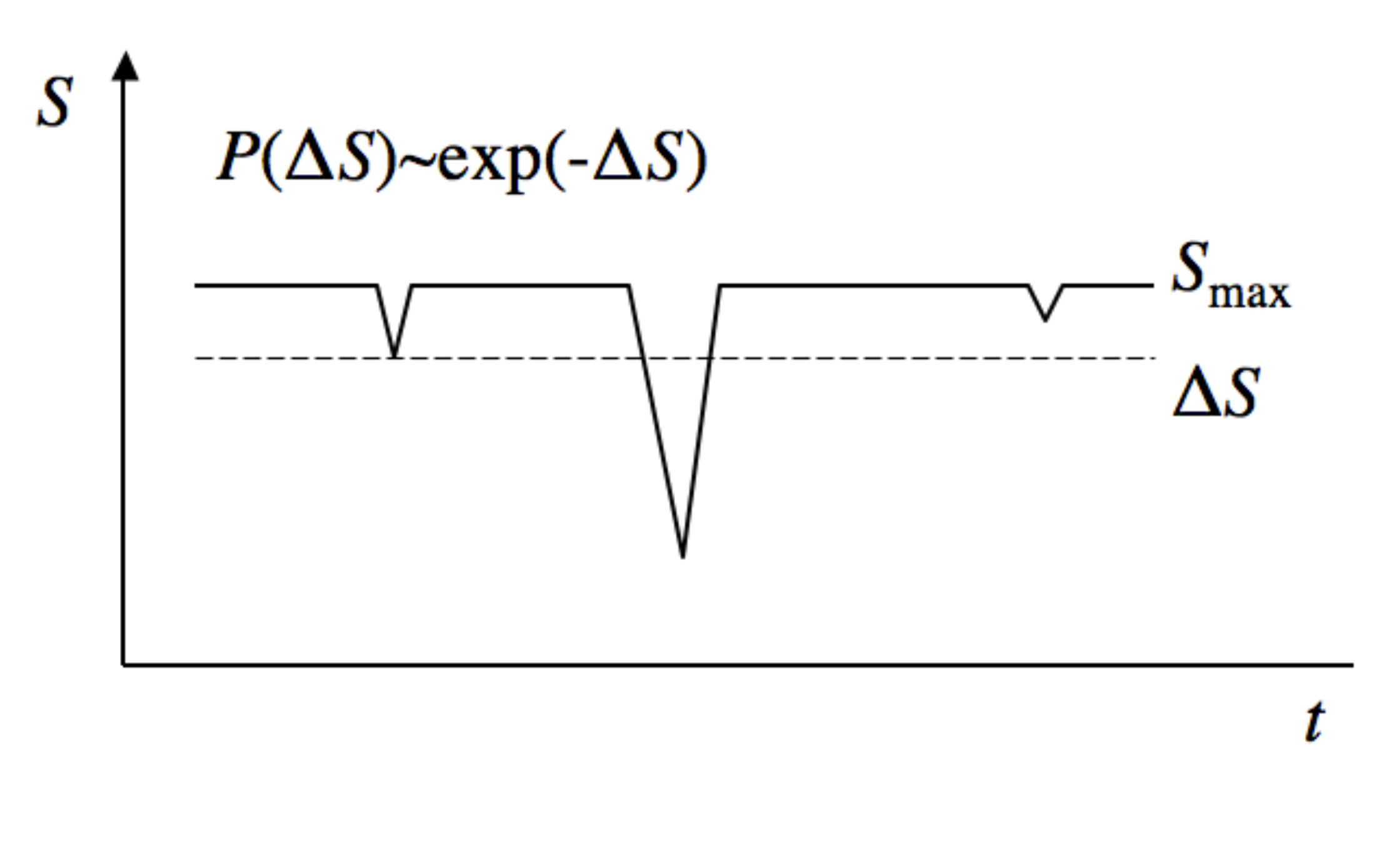}
\caption{If an isolated system has reached is maximum entropy, it can fluctuate spontaneously to a state of lower entropy. Larger fluctuations are statistically disfavoured with respect to smaller ones, so that for a given value of the entropy $S_{max}-\Delta S$ (dashed line) it is more likely that the entropy has already reached its minimum (like in the first trough) so that its evolution is time symmetric, rather than coming from a higher or lower entropy (like in the two points of the second trough). Applying this to our entire universe, we should not expect it to have even lower entropy in the past if its present situation were simply the result of a random fluctuation.}
\label{fig-2}       
\end{figure}

If an isolated system is already in its maximum entropy state it stays that way, so that it might be argued that macroscopically there is no time flow. In fact, this is not entirely accurate because you can have random fluctuations that result in a decrease of entropy: they are very unlikely (the probability goes as $P(\Delta S)\sim \exp(-\Delta S)$) but, if you wait long enough, your system can occasionally get into a more ordered state for a while (Fig.~\ref{fig-2}). A  consequence of this is that if one observes an isolated system to be initially far from equilibrium, one should not expect it to come from an even lower entropy state (which correspond to a smaller volume in phase-space and to an even lower probability) but from a higher entropy one, down from the maximum entropy. This is because of the exponential factor that suppresses large fluctuations.  

\subsection{Entropy in cosmology}
Let us try to apply this line of reasoning to the whole observable universe, considered as an approximately closed, isolated system. First of all, if we look at how the universe evolves, we see that there is a clear trend from simplicity/uniformity to complexity/lumpiness. We start with a very uniform gas in thermal equilibrium, and get more organized and more clumped, apparently moving very far from equilibrium as we move from past to future. This may seem to contradict what we know about the second law of thermodynamics. 

\begin{figure}
\centering
\includegraphics[width=0.8\textwidth]{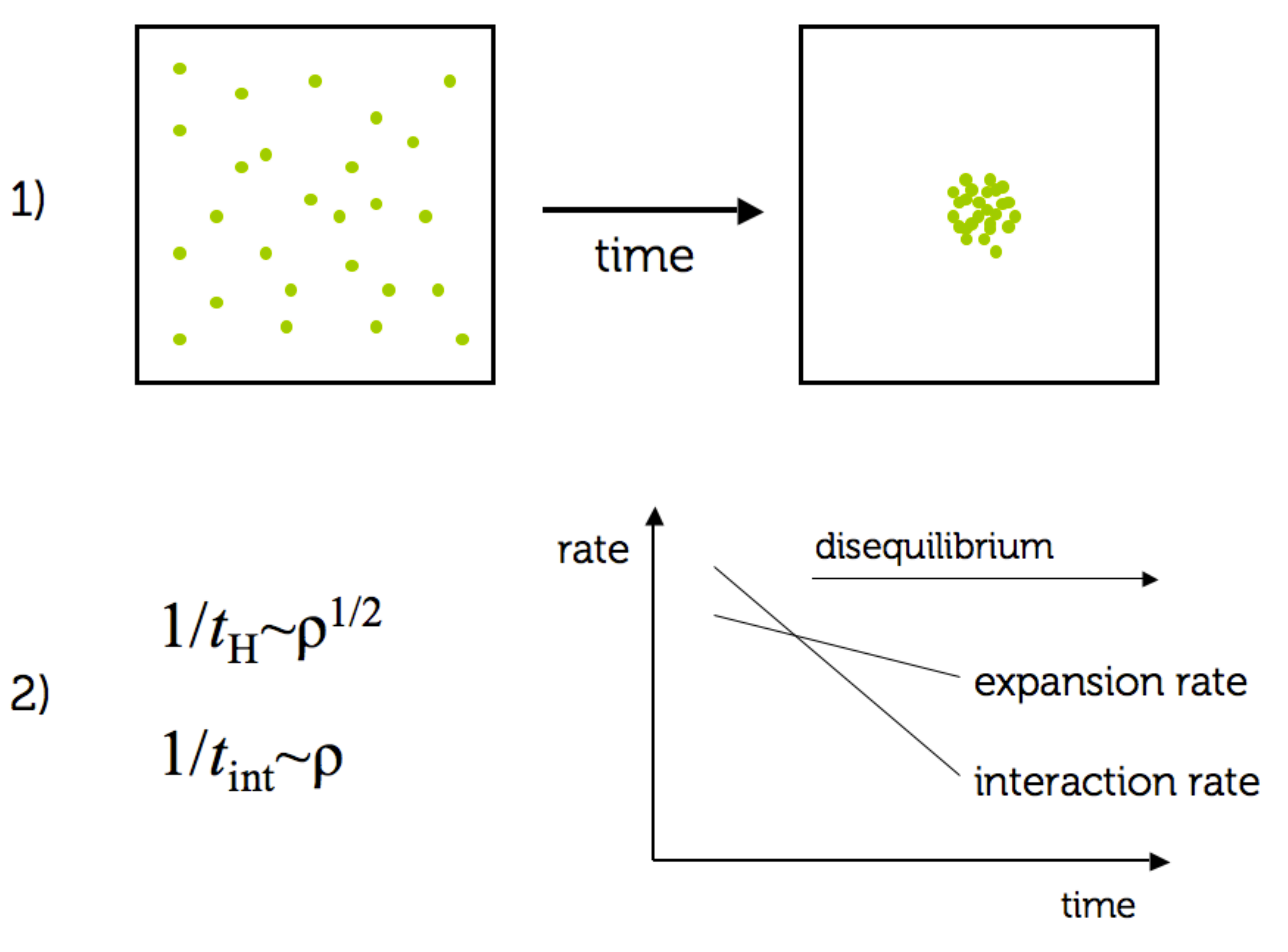}
\caption{When gravity is taken into account, matter can go from a more uniform state to a clumpier one: the universe becomes more structured and complex as time passes, apparently contradicting the second law. This is not the case as long as one takes into account the increase in entropy of the gravitational degrees of freedom. Another effect of gravity in an expanding universe is to eventually drive all physical interactions out of equilibrium, through the increase of the expansion rate.}
\label{fig-3}       
\end{figure}

The trick here is that we have to bring gravity into the business. Gravity does essentially two things (Fig.~\ref{fig-3}). First, it transforms an initially nearly uniform distribution into a clumpy one, going in the opposite direction of the above mentioned textbook example of particles in a box. So, the universe starts with tiny density fluctuations (very likely left from an inflationary phase) which are gravitationally amplified, resulting in clustered structures.  Second, gravity governs the way the cosmic clock ticks. As we mentioned above, one has to compare the Hubble time, $t_H\sim (G\rho)^{-1/2}$, with the time it takes for the various relevant physical processes to stay in equilibrium. The interaction timescale goes as $t_{int}\sim 1/\rho$: so, as the universe evolves, all processes naturally go out of equilibrium. This is, for example, the reason why life can support itself on a planet: energy can flow because there is a strong disequilibrium between the Sun temperature ($T\simeq 6000$ K), the surface temperature of Earth ($T\simeq 300$ K) and the temperature empty space ($T\simeq 2.7$ K). The idea is that there is a sink given by the gravitational degrees of freedom of the universe, where one can dump all the extra entropy created, and this gives the illusion that order is being created out of disorder---which is true, but only in branching systems that decouple from a state of increasing entropy \cite{reichenbach}: there is no violation of the second law, if one accounts correctly for the overall increase in entropy.

The problem is that we do not know exactly how to calculate entropy when gravity is put into the business. But we have reasonable estimates \cite{penrose}. If one lets a clump of isolated matter collapse into a black hole of area $A$, one gets the maximum entropy for that system, the Bekenstein-Hawking value $S_{BH}=A/4G\sim 10^{77} (M_{BH}/M_\odot)^2$. If one does the exercise with the entire mass of the universe one gets a number of order $S_{max}\sim 10^{121}$. This can get higher if one lets black holes evaporate, arriving finally to a de Sitter universe with entropy $S_{max}=3\pi/G\Lambda\sim  10^{123}$. In the past, before recombination, when the universe was smooth and dominated by radiation entropy, one gets a number of order $S_{CMB}\sim 10^{88}$. When inflation starts, this is even lower: $S_{inf}= 3\pi/G\Lambda_{inf} \sim 10^{12}$.  So, on one hand we see that there is an explanation for the arrow of time: the universe has not yet reached its maximum allowed entropy. On the other hand there is a  mystery on why the entropy was so low to start with. Inflation does not solve this, only makes things worse \cite{penrose}. 

\subsection{Time asymmetry in an expanding universe}
So we have a puzzle with time that is actually a cosmological puzzle. That the apparent time asymmetry was ultimately related to the properties of the cosmos was recognized very early: Boltzmann himself initially thought that the explanation was that the universe was eternal, and that its observed configuration was the result of an incredibly rare random fluctuation. This, however, was troublesome: if entropy had to fluctuate just enough in order to produce a universe with observers, it would have been much more likely to do so in much simpler situations (in the most extreme -- albeit seemingly preposterous -- scenario, it would be enough to have just one brain with fake memories floating in empty space, a so-called Boltzmann brain \cite{boltzmannbrain}). This is clearly not the case, since the universe shows a coherent evolution over the course of billions of years.   

Why, then, does the universe starts in such an incredibly low entropy state? No consensus has been reached yet, but we can give a few examples which are illustrative of different approaches. Some attempts go in the direction of leaving no freedom for the universe to choose its initial state, by assuming that there is some yet undiscovered law of nature that sets a low-entropy boundary condition in the past: the Weyl's curvature conjecture put forward by Penrose falls in this category, making the guess that the gravitational entropy can be expressed as some function of the Weyl tensor, and that the universe started in a smooth, ordered state with null Weyl curvature \cite{penrose1979}. Other explanations \cite{carrollchen} posit the existence a multiplicity of universes or \textit{multiverse}, as in the eternal inflation scenario, so that our big bang is just an occurrence in an infinite number of similar events: entropy can then increase without bound, and inflation can happen both towards the past and the future, so that everything is statistically time-symmetric on ultra-large scales. Within the context of inflation, others \cite{albrecht} have noted that there might be no deep explanation for why the universe starts in a very unlikely state, and that there must always be some ``special'' initial state which somehow resembles the ``rare fluctuation'' originally assumed by Boltzmann---although inflation has a way of explaining dynamically the evolution of a universe like ours from such an initial state in a more plausible way than its simple occurrence from a random fluctuation. Finally, some might argue that the low-entropy initial condition has to be taken as a brute fact, requiring no explanation (akin to ``why is there something rather than nothing?'' kind of questions).

\section{Conclusions}
I have outlined a few issues of standard cosmology which appear to have some direct or indirect connection to time. Whether or not they are pointing out to some deep insight into the nature of our universe is as yet unclear, but it is certainly worth pondering over.

\end{document}